\documentstyle[aps,preprint,prb,epsf]{revtex}
\title{The  Spectral Autocorrelation Function  in  Weakly
 Open  Chaotic Systems:  Indirect Photodissociation of Molecules}
\author{Y.  Alhassid$^1$  and Yan V. Fyodorov$^2$\cite{leave} }
\address{$^1$Center for Theoretical Physics, Sloane Physics
Laboratory, Yale University, New Haven, Connecticut 06520 \\
$^2$Fachbereich Physik, Universit\"at-GH Essen,
D-45117 Essen, Germany }
\begin{document}
\draft
\maketitle

\begin{abstract}

 We derive the statistical limit of the spectral autocorrelation function and 
of the survival probability for the indirect photodissociation of molecules 
in the regime of non-overlapping  resonances. The results are derived in the
 framework of random matrix theory, and hold more generally for any
chaotic quantum system that is weakly coupled to the continuum.
  The ``correlation hole''  that characterizes the spectral autocorrelation in 
 the bound molecule diminishes as the typical average total  width of  a 
 resonance  increases.
 
\end{abstract}


\newpage

 Quantum systems that are classically chaotic are believed to exhibit
 statistical fluctuations in their spectra and wavefunctions that 
are universal.\cite{Bo}  These universal properties are well-reproduced by
 the assumption  that the Hamiltonians belong to an ensemble of 
 Hamiltonians that are consistent with  the underlying symmetries,
 but are otherwise random.  Such random Hamiltonians
 are described by random matrix theory (RMT),\cite{RMT,Mehta} 
and lead to level repulsion, long range  correlations in the spectrum,
 and Porter-Thomas statistics for the transition intensities. 
RMT was initially developed to explain the statistical  properties
 of the neutron resonances in the compound nucleus.\cite{RMT} 
 However, similar 
statistical fluctuations are observed in a variety of molecular systems.
These include fluorescence excitation spectra of polyatomic molecules, 
e.g., 
 the  vibrational levels and intensities of nitrogen dioxide 
\cite{ZK88,GD95} and pyrazine. \cite{Le88}  
 Level repulsion  was  observed in acetylene, \cite{Ab86} and
 statistical signatures of RMT were found in the Stark level-crossing spectra 
of  formaldehyde at the dissociation threshold.  \cite{WM90a} 
 At low excitations the molecular Hamiltonian is
approximately separable (though anharmonic), and the vibrational  eigenstates
 are described by normal modes, allowing the assignment of good quantum
  numbers (``regular'' regime). However, at higher excitations the normal modes
 are strongly mixed and  the spectrum is better characterized  in terms of its 
statistical properties  (``chaotic'' regime).  A transition from regular to 
chaotic spectrum  was  observed in the vibrational  levels of CS$_2$.\cite{Pi91}

Statistical  measures of a spectrum include  the nearest neighbors level 
spacing distribution and spectral rigidity  (e.g. the  Mehta-Dyson $\Delta_3$ 
statistics). 
Such measures  require accurate experimental determination of a complete set 
of levels.\cite{GD95}
 In practice this is not always possible  since the finite experimental 
resolution makes it difficult to determine the statistics of the small level 
spacings. In molecular spectroscopy it is particularly difficult to measure a 
complete set of levels because  the  density of states is very large at higher
 excitations. 

 A measure of  chaos that is less sensitive to experimental resolution
  is the spectral autocorrelation function and its counterpart in the time
 domain, the survival probability. \cite{P82,LK91} 
Since the spectral autocorrelation
 involves the convolution of the spectrum (i.e. strength function) with itself,
 it is less sensitive to experimental noise. 
 The spectral  autocorrelation function 
 is easily constructed from the experimental  measurements. 
For example,  using lasers one can excite  molecules that are 
vibrationally and rotationally cooled, and 
measure their absorption or  fluorescence spectra 
(i.e., the strength function of the dipole operator). 
The convolution of the measured spectrum with itself  is the
 spectral autocorrelator or the power spectrum. 
 The corresponding quantity in the time domain, i.e., the Fourier transform 
of the power spectrum,
is  the probability that an  initially prepared non-stationary state 
(e.g.,  the state generated by the operation of the dipole operator  on the
 ground state of the molecule)  remains in its initial state at a later time.
 For a chaotic or complex 
system, it was suggested  that the power spectrum  and the associated 
survival probability of the experimentally prepared state
 are characterized by the existence of a  ``correlation
 hole''  (i.e. a  pronounced local minimum) that originates in  the repulsion 
of energy levels. \cite{Lev86}  A correlation hole was observed  in the 
spectrum of highly excited vibrational levels of complex molecules like
 methylglyoxal \cite{Lev86} and  acetylene. \cite{Pi87}
The correlation hole is a signature of  chaos and it disappears 
 when the dynamics of the system becomes regular.

 The average  survival probability and power spectrum  in chaotic systems
were evaluated in a closed form using RMT for the various classes of 
 Gaussian ensembles, \cite{AL92} and in the framework of 
a scattering model.\cite{GW90}  Similar expressions can be 
obtained by considering discrete time evolution for the
 circular Dyson ensembles. \cite{LI97}   The survival probability
 (and the associated power spectrum)  carries information on both spectral 
 and eigenvector statistics. \cite{LK91}    
The derivation was restricted to systems that are closed and have discrete 
spectra. However, in many physical situations we are interested in the
 spectrum at high excitations, where the eigenstates are coupled to the 
continuum and become resonances.  
 The coupling of the quantum system to the
 continuum is expected to diminish the correlation hole.  Experimentally
 one usually observes a quenched correlation hole, an effect attributed
 to unobserved good quantum numbers \cite{Pi87} and/or the transition
 to regular dynamics. \cite{WB91,AW93}
  However, since the correlation hole can also be affected by the 
 coupling of the closed system to its environment, it is important to derive
  the correlation hole quantitatively  in this more general situation.  

 An important class of processes where coupling to the continuum is important
 is the photodissociation  of molecules.   In particular we are
 interested in  indirect photodissociation which proceeds through resonances
 of the excited molecular complex. \cite{Sc93}  A light pulse excites the 
molecule to 
a higher electronic state  at energies above the dissociation threshold, 
but where a potential barrier hinders the immediate dissociation of the
 molecule.  The barrier is usually formed because of  the avoided crossing of 
two diabatic electronic potential surfaces (one that is binding and one that is
 repulsive).  Vibrational excitations (below and above the barrier) become
 resonances whose widths depend on the coupling between the binding
and repulsive manifolds.  Such resonances can also be populated in molecular 
reactions that proceed through an excited molecular complex. 

 Recently, the  universal correlator for the bound-to-continuum 
 strength function  was derived  for any degree 
of openness of the quantum system, \cite{FA98}  by using the method of the
 supersymmetric non-linear  $\sigma$ model.\cite{Efrev,VWZ}
However, the general expression  turned out to be rather 
complicated. Here we show that  in the regime of isolated resonances
 (i.e. weak coupling) a much simpler closed form  can be derived 
 using random matrix  techniques. This regime of non-overlapping resonances
corresponds to  indirect photodissociation  at energies where the corresponding 
vibrational resonances
 lie below the barrier in the excited electronic potential surface. The width 
of such resonances is smaller than a typical separation between vibrational 
states  by a factor of  up to several thousand. 
 The derivation in this paper is  accomplished in the
 time domain  where the survival probability of the experimentally prepared 
non-stationary state is averaged over the ensemble. This derivation is
 analogous to that used in Ref. \onlinecite{AL92}  for closed systems.
However, in the case of  an open quantum system, it is necessary to take into
 account the decay of the system into the continuum.  The probability for 
an open chaotic system (that is initially bound) to remain bound at a later
 time was calculated in Ref. \onlinecite{HDM}.   The coupling of the system to
 the continuum is described by the average partial widths of the system to decay
 into each of the open dissociation channels. The spectral correlator 
derived in this paper in the quantum ergodic limit  (see Eqs.  (\ref{corr1}) 
and (\ref{func})
below) depends only on  these average
 partial ``dissociation'' widths  (and the symmetry class of the Hamiltonian), 
 but is otherwise universal.  We find that the correlation hole diminishes
 when the partial
 widths of the resonances become larger or when the number of open channels
 increases.

 An open system  is described
 by an effective Hamiltonian ${\cal H}_{eff} = H - i \Gamma/2$, 
where $H$ is the 
Hamiltonian of  the system when  it is uncoupled from the continuum, 
and $\Gamma$ is a matrix describing the coupling of the system to
 external open channels. \cite{VWZ,WM90b,FS}  In indirect dissociation,
$H$ represents that part of the Hamiltonian that includes 
 ``binding'' potential energy surface only, while the open channels describe
 the possible dissociative states of the molecule.
 The eigenstates  $\mid n \rangle$ of  ${\cal H}_{eff}$
 have complex eigenvalues $E_n - i \Gamma_n/2$, describing   resonances
 with energies $E_n$ and widths $\Gamma_n$.  
  Bound-to-continuum transitions that proceed through 
resonance states are described by  the strength function of
 the corresponding transition operator $T$ \cite{Sc93}
 \begin{eqnarray}\label{strength}
   \sigma (E)  = {\sum_n} \mid \langle n \mid T \mid g \rangle \mid^2 
     \frac{\Gamma_n/2}{(E-E_n)^2 + \Gamma_n^2/4} \;.
\end{eqnarray}
In Eq. (\ref{strength})  each transition intensity 
(to a state imbedded in the continuum) is weighted
by  a Lorentzian centered at  the  resonance energy.
In indirect photodissociation, the photoabsorption spectrum 
is proportional to Eq. (\ref{strength}) where $T$ is the dipole operator 
of the molecule. In the following we consider only weakly coupled systems that 
are characterized by  non-overlapping and narrow resonances (relative to their 
mean spacing). In this case $\mid n \rangle$
 are just  the eigenstates of the closed system Hamiltonian $H$, 
$E_n$ are the eigenvalues of $H$ (to leading order we ignore shifts in 
eigenvalues), and the  widths are given by
\begin{eqnarray}\label{width}
      \Gamma_n = \sum_{n=1}^\Lambda \Gamma_{nc}  \;,
\end{eqnarray}
where  $\Gamma_{nc} =  \mid \gamma_{nc}\mid^2$ is the partial width 
of a resonance level $n$ to decay into an open channel $c$ ($\gamma_{nc}$
 is the partial width amplitude). We assume that there are $\Lambda$
 open channels.
In the $R$-matrix  formulation of reaction theory it is possible  to express
 the partial width amplitude as an overlap integral of the eigenstate  $n$
 and the corresponding
 channel wavefunction at the interface  of the internal ``interaction''
region and the external  asymptotic region. \cite{LT58}

The total strength satisfies the sum rule  
\begin{eqnarray}\label{sum}
\int \sigma(E) dE = \sum_n \mid \langle n \mid T
 \mid g \rangle \mid^2 = \langle g \mid T^\dagger T \mid g \rangle \;,
\end{eqnarray}
so that the experimentally prepared state   $T | g\rangle$ can be normalized
 by defining    $| \alpha \rangle\equiv T  | g \rangle 
/ \sqrt{\langle g \mid T^\dagger T \mid g \rangle}$. 
The normalized strength function $\tilde{\sigma}(E) = \sigma (E)/
\int  \sigma (E^\prime dE^\prime$
  can then be rewritten as
\begin{eqnarray}\label{strength1}
   \tilde{\sigma} (E) = {\sum_n} \mid \langle n \mid \alpha \rangle \mid^2
    \frac{1}{\pi} \frac{\Gamma_n/2}{(E-E_n)^2+\Gamma_n^2/2} \;.
\end{eqnarray}
 In the following we shall  simply use $\sigma(E)$ to denote the normalized
 strength function.

  The spectral autocorrelation function $G(\omega)$ is defined in
 terms of the strength function by \cite{LK91}
\begin{eqnarray}\label{spect}
      G(\omega) =  \int_{-\infty}^{\infty}  \sigma( E)  \sigma (E+ \omega) dE
  \;.
 \end{eqnarray}
In the time domain we can define $C(t)$ as the Fourier transform of
 $\sigma (E)$, $C(t) =  \int_{-\infty}^{\infty}  \exp(-i E t)  \sigma (E) dE$.  
Using (\ref{strength1})  we find
\begin{eqnarray}\label{amp}
C(t) =  \sum_n |\langle n | \alpha \rangle|^2 e^{-i E_n t - \Gamma_n |t|/2} =
 \langle \alpha \mid \alpha(t) \rangle \;,
\end{eqnarray}
where $ \mid \alpha (t) \rangle \equiv \exp(-i {\cal H}_{eff} t)\mid
 \alpha \rangle$ is
 the state at time $t$ that evolves  from the  experimentally prepared 
 non-stationary initial state $\mid \alpha \rangle$ under the effective
 Hamiltonian ${\cal H}_{eff}$.  
Thus the Fourier transform of  the spectral autocorrelation $G(\omega)$ is
just the survival probability   $P(t) = | C(t) |^2$, i.e. 
the probability at time $t$ that the system remains at its initially prepared 
state. Notice that for an open system, decay (into the continuum) is included
 in the time evolution.  Using (\ref{amp}), the survival probability can be
 expressed as \begin{eqnarray}\label{prob}
   P(t) = {\sum_n}\mid \langle n \mid \alpha\rangle \mid^4 
 e^{-\Gamma_n \mid t \mid}
+ \sum_{n \neq m} \mid \langle n \mid \alpha \rangle
   \mid^2 \mid \langle m \mid \alpha \rangle \mid^2  e^{i(E_n-E_m)t}
   e^{{-\Gamma_n + \Gamma_m \over 2} \mid t \mid }  \;.
\end{eqnarray}
We would like to calculate the ensemble average of $P(t)$. In RMT,  the
 eigenvalues and eigenvectors are statistically independent. This implies
 that $\mid \langle n \mid \alpha\rangle \mid^2$ and $\Gamma_n$ (which
 are determined by the eigefunctions)  are statistically independent from 
the spectrum $E_n$. Moreover, we assume that the prepared state 
$\mid \alpha \rangle$  has no overlap with the channels, so that the
 projection of  the eigenstate
$\mid n \rangle$ on $\mid \alpha \rangle$ and on the channels (i.e. partial
 width amplitudes) are also statistically independent. We conclude that all 
three quantities 
$|\langle n | \alpha\rangle |^2$, $\Gamma_n$ and $E_n$, 
appearing in Eq. (\ref{prob}), are statistically independent. Consequently,
 the ensemble average on the right hand side of (\ref{prob}) can be taken
 over each factor separately.  In RMT, the eigenvector components are 
distributed randomly on the $N$-dimensional sphere, and thus
 one finds for a fixed normalized vector $|\alpha\rangle$:
\begin{eqnarray}\label{comp}
 {\overline{\mid\langle n \mid \alpha \rangle\mid^2\mid
  \langle m \mid \alpha\rangle \mid^2} \over \overline{\mid
 \langle n\mid\alpha\rangle \mid^4}}
      & =  & \frac{\beta}{\beta +2} \;;\\ \nonumber
    N \overline{\mid \langle n \mid \alpha \rangle \mid^4}
 & = & \frac{\beta +2}{N\beta +2} \;.
\end{eqnarray}
Here $\beta$ is a constant that depends on the corresponding Gaussian ensemble.
 For systems that conserve  time-reversal symmetry the relevant ensemble is
 the Gaussian  orthogonal ensemble (GOE) and $\beta=1$, while  systems
 that break time-reversal symmetry are characterized by the Gaussian unitary
 ensemble (GUE) and $\beta=2$.   Chaotic nuclear  and molecular
 systems always have $\beta=1$. 
The partial width amplitudes can be shown to have a joint 
Gaussian distribution \cite{AL95} 
\begin{eqnarray}\label{joint}
     P(\gamma) \propto (det M)^{-\frac{\beta}{2}}
           e^{-\frac{\beta}{2} \gamma^\dagger  M^{-1} \gamma} \;,
\end{eqnarray}
 where    $M_{cc^\prime} \equiv \overline{ \gamma^\ast_c\gamma_{c^\prime}}$ 
is the channel correlation matrix.  Eq. (\ref{joint}) follows from RMT
 using the fact that the partial widths can be described as the
 projection of the internal wavefunction (of the closed system)
 on the open channels.  In the following we assume, without loss of 
generality,  that  all  the $\Lambda$ channels are uncorrelated.  Otherwise, 
it is always possible to transform to the set of eigenchannels. 
The eigenvalues of $M$ are 
just the average partial widths $\bar{\Gamma}_c$ of the eigenchannels. 
Using $\Gamma_n = \sum_{c=1}^\Lambda  \mid \gamma_{nc}\mid^2$ and the
 known distribution (\ref{joint}) of the partial width amplitudes 
$\gamma_{nc}$, it is possible to derive
the total width distribution $P(\Gamma_n)$. \cite{WM90b,AL95}
Such distributions  were observed,  for example, in  the  measured
decay rates of  highly excited vibrational states of formaldehyde 
in its ground electronic states $S_0$.\cite{WM90a,WM90b}
They were also reproduced in numerical 
simulations of a two-dimensional model chaotic cavity. \cite{AL95}
 For the ensemble average of the decay factor in Eq. (\ref{prob}) we then find
\begin{eqnarray}\label{expg}
     \overline{e^{-\frac{\Gamma_n}{2}\mid t\mid}} =
        \det \left( 1 + \frac{\mid t\mid}{\beta}  M \right)^{-\beta/2} =
 \prod_c {1 \over \left(1+\frac{\mid t\mid}{\beta}
 \bar{\Gamma}_c\right)^{\beta/2}} \;.
\end{eqnarray}

To leading order in $1/N$, $\Gamma_n$ and $\Gamma_m$ ($n \neq m$)
are statistically independent so that  $\overline{e^{-\frac{\Gamma_n
 + \Gamma_m}{2}
\mid t\mid} } = \left( \overline{e^{-\frac{\Gamma_n}{2} \mid t\mid}} 
\right)^2$. From the known RMT spectral statistics we also have  
\begin{eqnarray}\label{expe}
 (N-1) \overline{e^{i(E_n-E_m)t}} = \delta \left( {\frac{t}{t_H}}\right)
             - b_{2,\beta}\left( {\frac{t}{t_H}} \right) \;,
  \end{eqnarray}
where   $t_H \equiv 2\pi \bar{\rho}$ ($\bar{\rho}$ is the average 
level density) is the Heisenberg time.
$b_{2,\beta}(\tau)$  is the two-level form factor, defined as the
 Fourier transform of the two-point cluster function $Y_{2,\beta}(s)$. 
The form factor is known
 analytically for both the GOE and GUE ensembles \cite{Mehta}
\begin{eqnarray}\label{form}
  b_{2,\beta} (\tau) = \left\{ 
      \begin{array}{crc} \left[1-2 |\tau| +|\tau| \ln(1+2|\tau|) \right]  
\theta(1- |\tau| ) + \left[ |\tau| \ln\left({2|\tau|+1 \over 2|\tau | -1 }
\right) - 1 \right] \theta(|\tau|-1)  &\;\;\;& (\beta =1)  \\
            (1- |\tau|)  \theta(1- |\tau| )  &\;\;\;& (\beta =2)      
 \end{array}  \right.
\end{eqnarray}

 Combining (\ref{comp}), (\ref{expg}) and (\ref{expe}), and measuring
 time in units of the Heisenberg time ($\tau = t/t_H$), we find for the 
ensemble average of the  survival probability
\begin{eqnarray}\label{surv}
  N \overline{P(\tau)}   =  
 \delta(\tau)
  -  b_{2, \beta}(\tau)  \prod_c 
  { 1 \over \left(1+ \mid \tau \mid  T_c/\beta \right)^\beta}
+ \frac{\beta +2}{\beta}
 \prod_c   { 1 \over \left(1+  2\mid \tau \mid  T_c/\beta \right)^{\beta/2}} \;,
\end{eqnarray}
where we have used the transmission coefficients $T_c$. 
  For a weakly coupled system, the average partial width $\bar{\Gamma}_c$
 is related to the transmission coefficient $T_c$ in channel $c$ by
 $\bar{\Gamma}_c = T_c/2\pi \bar{\rho}$. \cite{VWZ}
 In the limit of closed system $T_c=0$ for all channels and
 Eq. (\ref{surv}) reduces to  Eq. (12)  of Ref. \onlinecite{AL92}.

The ensemble average of the spectral autocorrelation  $G(\omega)$ is 
found by  taking the inverse Fourier transform of  Eq. (\ref{surv}). 
   Defining  the reduced spectral autocorrelation function 
 $g(\omega)  \equiv \left[ \overline{\sigma (E) \sigma (E+\omega)} - 
\overline{\sigma (E)}^2 \right ]/
  \overline{\sigma(E)}^2 = \left[ (\bar{\rho}/ N) \overline{G(\omega)} - 
\overline{\sigma(E)}^2 \right]/
\overline{\sigma(E)}^2 $, we find the following universal form   
\begin{eqnarray}\label{corr}
 g_\beta(\omega) = 2 \left[ \frac{\beta +2}{\beta} \int^\infty_{0} d\tau
  \cos(2\pi \omega\tau) \prod_c 
   {1 \over \left (1+\frac{2}{\beta} T_c\tau \right)^{\beta/2}}
  - \int^\infty_{0} d\tau \cos(2\pi \omega \tau) b_{2 \beta}(\tau)
    \prod_c {1 \over \left(1+\frac{1}{\beta} T_c \tau\right)^\beta} \right] \;.
\end{eqnarray}
  By defining a function $F_{\beta}(\omega)$ in the energy domain 
\begin{eqnarray}\label{func}
F_\beta(\omega) = 2 \int^\infty_{0} d\tau
  \cos(2\pi \omega\tau) \prod_c 
   {1 \over \left (1+\frac{2}{\beta} T_c\tau \right)^{\beta/2}} \;,
\end{eqnarray}
we can rewrite the reduced spectral autocorrelation  in the form 
\begin{eqnarray}\label{corr1}
g_\beta(\omega) =  \frac{\beta +2}{\beta} F_\beta(\omega)
  - \int_{-\infty}^\infty d \tilde{\omega} Y_{2,\beta}(\tilde{\omega}) 
F_{2 \beta}
 (\omega - \tilde{\omega}) \;.
\end{eqnarray}

 There are two interesting limits for Eq. (\ref{corr1}). In the limit of a 
closed system all  transmission coefficients vanish  ($T_c=0$),
 and Eq. (\ref{corr}) reduces to \cite{AL92}
\begin{eqnarray}\label{corr_c}
   g_\beta(\omega) = \frac{\beta+2}{\beta} \delta (\omega) - Y_{2,\beta} 
(\omega) \;.
\end{eqnarray}
 The second term in (\ref{corr_c})  gives rise to the correlation
 hole discussed in the introduction. Another  limit  for $\Lambda$ 
equivalent open channels  ($T_c=T$ for all $c$) 
 is the limit  $\Lambda \to \infty$ such that $\Lambda T =\kappa$ is kept 
constant. This is equivalent to taking the limit of a large number of equivalent
 channels while keeping the total average width $\bar{\Gamma}$
 of the level constant. In this limit the integrand in (\ref{corr}) becomes
 $e^{- \kappa |\tau|}$ and $F_\beta(\tau) = (\kappa/2) \left[ (\pi \omega)^2
 + (\kappa/2)^2 \right]^{-1}$ is a Lorentzian which is independent 
of $\beta$. Thus we obtain \cite{FA98}
\begin{eqnarray}
g_\beta(\omega) =  \frac{\beta +2}{\beta}  {\kappa/2 \over  (\pi \omega)^2
 + (\kappa/2)^2 }  - \int_{-\infty}^\infty d \tilde{\omega} 
Y_{2,\beta}(\tilde{\omega})  { \kappa/2 \over
\pi^2  (\omega - \tilde{\omega})^2 + (\kappa/2)^2} \;.
\end{eqnarray}

\begin{figure}

\epsfxsize= 12  cm
\centerline{\epsffile{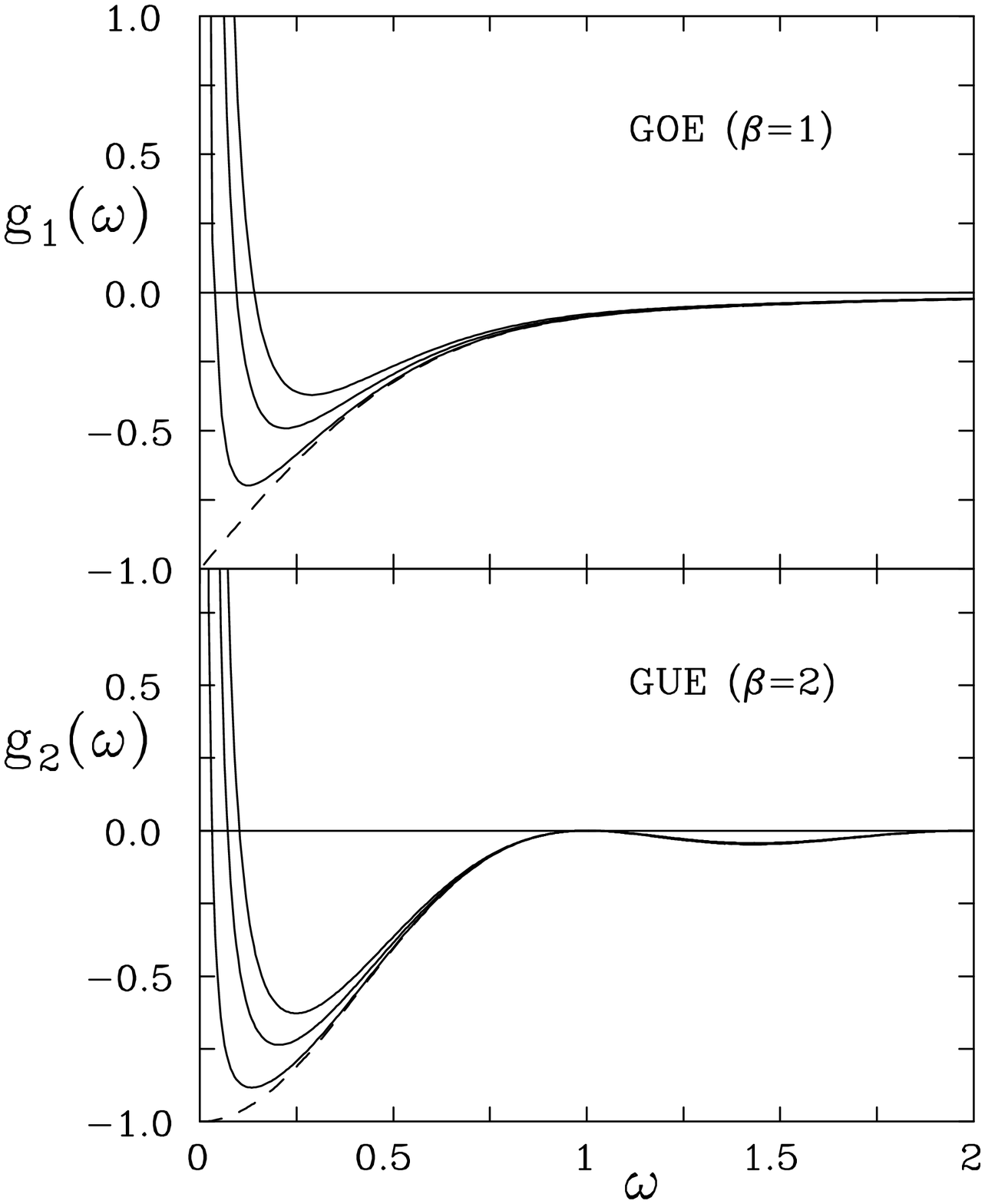}}

\vspace{1 cm}

\caption
{ The reduced spectral autocorrelation function $g(\omega)$ 
 vs. $\omega$ for  $\Lambda=1,5$ and $10$ equivalent channels with $T=0.01$
 (solid lines).  The minimum of $g(\omega)$ (``correlation hole'') 
 becomes shallower as the number of channels increases.
For reference we also show by dashed line  $g(\omega)$ for a
 closed system (without the $\delta$ function at $\omega=0$). 
Top: GOE statistics ($\beta=1$); bottom: GUE statistics ($\beta=2$).}
\label{fig1}

\end{figure}

 Eqs. (\ref{corr1}) and (\ref{func}) provide a closed expression for
 the universal spectral  autocorrelation
 in  weakly open chaotic systems, and  is the main result of this paper.
In Fig. \ref{fig1}  we compare the GOE and GUE  reduced  spectral  
autocorrelators for an open system coupled to $\Lambda = 1, 5$ and $10$ 
equivalent
 channels with a partial transmission coefficient of $T=0.01$ in each channel.
For  reference, we show by a dashed line the  spectral  correlator for 
a closed system (not including the $\delta$ function at $\omega=0$).
 As the number of channels increases, the correlation hole shrinks.  
It is seen that for the same number of open channels,  the GOE
 correlation hole is shallower than the GUE hole.
Thus the GOE hole disappears faster when opening  the system  
   to more channels. This could be explained by the stronger level repulsion in
 the GUE case.  A similar effect  (not shown here) is observed 
for a fixed number of channels as the transmission coefficient increases. 
Generally, we observe a shallower hole as the total mean resonance
 width increases. \cite{FA98}

  In conclusion, we have derived in closed form the spectral 
autocorrelation function and survival probability for bound-to-continuum 
transitions that proceed through a regime of isolated resonances.  The 
correlator is 
universal and  depends only on the symmetry class (i.e. conserved or broken
 time-reversal symmetry),  and the partial dissociation widths (or transmission
 coefficients) for the open channels.  
The correlation hole, a signature of chaos in both the power spectrum and
 survival probability, disappears gradually as more channels are opened, 
and it does so faster in systems that  conserve time-reversal symmetry.
It would be interesting to compare our prediction for the correlation hole
in open systems  (for the case of time-reversal symmetry) 
with actual measurements or numerical simulations   
of indirect  photodissociation  spectra of complex molecules.

\end{document}